# Polar self-organization of ferroelectric nematic liquid crystal molecules on atomically flat Au(111) surface

Alexandr A. Marchenko[1,2], Oleksiy L. Kapitanchuk[3], Yaroslava Yu. Lopatina[1,2], Kostiantyn G. Nazarenko[4,5], Anton I. Senenko[1], Nathalie Katsonis[2], Vassili G. Nazarenko[1,6], and Oleg D. Lavrentovich[7]

[1] *Institute of Physics of the National Academy of Sciences of Ukraine, 46 Nauki ave., Kyiv, 03028, Ukraine*

[2] *Stratingh Institute for Chemistry, University of Groningen, Nijenborgh 4, 9747AG Groningen, The Netherlands*

[3] *Bogolyubov Institute for Theoretical Physics of the National Academy of Sciences of Ukraine, 14-B Metrologichna str., Kyiv, 03143, Ukraine*

[4] *Institute of Organic Chemistry of the National Academy of Sciences of Ukraine, Kyiv, Ukraine;*

[5] *Enamine Ltd, Kyiv, Ukraine*

[6] *Institute of Physical Chemistry, PAS, Kasprzaka 44/52, 01-224 Warsaw, Poland*

[7] *Advanced Materials and Liquid Crystal Institute, Department of Physics, Materials Science Graduate Program, Kent State University, Kent, OH 44242, USA.*

## Abstract

Understanding nanoscale mechanisms responsible for the recently discovered ferroelectric nematics can be helped by direct visualization of self-assembly of strongly polar molecules. Here we report on scanning tunneling microscopy (STM) studies of monomolecular layers of a ferroelectric nematic liquid crystal on a reconstructed Au(111) surface. The monolayers are obtained by deposition from a solution at ambient conditions. The adsorbed ferroelectric nematic molecules self-assemble into regular rows with tilted orientation, resembling a layered structure of a smectic C. Remarkably, each molecular dipole in this architecture is oriented along the same direction giving rise to polar ferroelectric ordering.

**Introduction**

The simplest example of an orientationally ordered liquid is a uniaxial paraelectric nematic (N) liquid crystal (LC). The average orientation of molecules in the N phase is described by a unit vector $\hat{\mathbf{n}}$, called the director. The states $\hat{\mathbf{n}}$ and $-\hat{\mathbf{n}}$ are equivalent since the order is apolar, with no distinction between heads and ends of the rod-like molecules [1]. Recent discoveries [2-6] revealed a new ferroelectric nematic ($N_F$) phase with the polar orientation of molecules carrying large dipole moments, about 9-10 D, which is 1.5-2 times stronger than the dipole moments of many conventional N mesogens. Unidirectional alignment of dipoles yields spontaneous macroscopic electric polarization **P**, locally parallel to the director $\hat{\mathbf{n}} \equiv -\hat{\mathbf{n}}$. The large polarization enables strong polarity-sensitive linear coupling to an applied electric field. Molecular orientation in an $N_F$ slab could be realigned by electric fields on the order of $\sim 10^2$ V/m, thousand times smaller than those used to reorient conventional Ns [6].

The observation of $N_F$ has been followed by the discovery of an antiferroelectric smectic $SmZ_A$ with **P** parallel to equidistant layers [7] and ferroelectric smectic A ($SmA_F$), in which the molecules form layers perpendicular to **P** [8,9]. Besides stronger dipole moments, the $N_F$, $SmZ_A$, and $SmA_F$ molecules are otherwise similar to the molecules forming the paraelectric N. The $N_F$, $SmZ_A$, and $SmA_F$ phases readily transform into the N upon minute changes in temperature or as a result of a single-atom replacement in the molecules. The nanoscale underpinning of this delicate balance is far from being understood. The picture is certainly far more complex than a simple electrostatic concept that two parallel dipoles attract when placed head-to-tail and repel when placed side-by-side. Direct experimental visualization of self-organization of polar molecules could provide insight into the mechanisms. This work presents such a visualization, enabled by scanning tunneling microscopy (STM) of two-dimensional monolayers at atomically flat crystal substrates, Fig.1.

Interactions with bounding substrates are known to affect ferroelectric nematics strongly [6,10,11]. The polarization **P** could be aligned by confining the material between two glass plates with unidirectionally rubbed polymer coatings [5-19]. Alignment of the $N_F$ at solid crystalline substrates is studied much less. It is not clear whether **P** will be aligned by a crystal substrate with well-defined crystallographic axes but no in-plane polar direction.

To visualize ferroelectric nematic molecules, we explore self-assembled monolayers (SAMs) of the $N_F$ material 2,3′,4′,5′-tetrafluorobiphenyl-4yl 2,6-difluoro-4-(5-propyl-1,3-dioxan-2-yl)benzoate (also known as DIO [2]), Fig.1b (inset), adsorbed at an atomically flat reconstructed Au (111) substrate. STM reveals the polar orientation of the $N_F$ molecules in the monolayer. These arrangements are in striking contrast to non-polar alignment in monolayers of the N mesogen 4-n-decyl-4'-cyanobiphenyl (10CB), Fig.1d (inset), at the same experimental conditions.

**Materials and methods**

The reconstructed Au(111) substrates are prepared from gold films deposited in ultra-high vacuum (~$5 \times 10^{-8}$ Pa) onto a freshly cleaved mica heated to ~600K, followed by annealing in a propane-air flame. On cooling from the isotropic (I) phase, the phase sequence of DIO is I−174°C −N−82°C −SmZ$_A$−66°C −N$_F$−34°C −Crystal, where SmZ$_A$ is an antiferroelectric smectic with a partial splay [2,9,20]. At solid substrates, the DIO molecules and thus **P** align parallel to the substrate, to avoid a strong surface charge [10,18,21-26]. DIO is dissolved in tetradecane (99% purity, Aldrich), to a concentration of approximately ~0.05 mg/ml. The droplet of the DIO solution is placed onto the Au(111) substrate. Atomic-level imaging is performed with a PicoSPM from Molecular Imaging (USA) at ambient conditions. The STM tip is a Pt/Ir (80:20) wire sharpened by mechanical cutting. Typical imaging conditions correspond to a bias voltage $U_t$ in the range 50-500 mV and a tunneling current $I_t = 50\text{-}500$ pA. The error of measured distances is within 5%. Several STM images in the constant-current mode are recorded with different samples and tips to verify reproducibility. Film preparation and STM imaging are performed at room temperature (~20°C).

**Results and discussion**

Figure 1a shows the large-scale STM image of the $N_F$ film at the Au(111) substrate at ambient conditions. The molecules form well-ordered bright layers. The layers tend to be perpendicular to the atomic step edges of the Au(111) substrate, as seen in the top right corner of Fig. 1a. Since the step edges of Au(111) surface are parallel to the crystallographic <110> direction [27,28] one concludes that molecular layers are stacked along the perpendicular <112> direction. The layers in neighboring domains make angles 60°±5° or 120°±5° with each other that correspond to the angle between the equivalent crystallographic <110>, <101> and <011> directions of Au(111) surface.

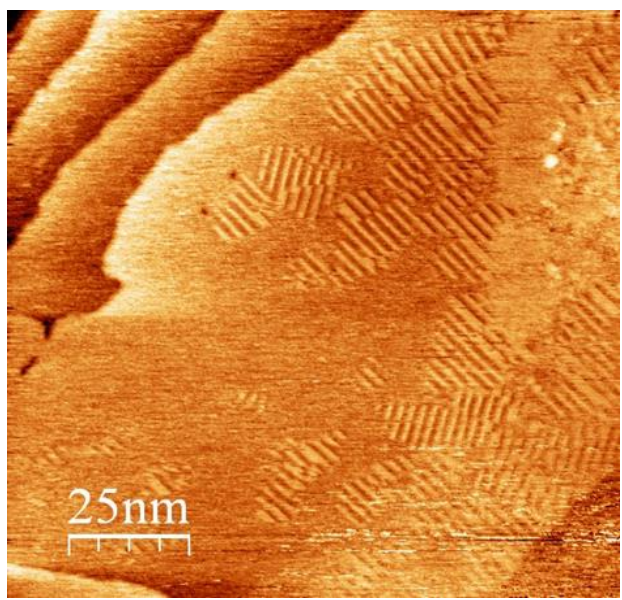

a

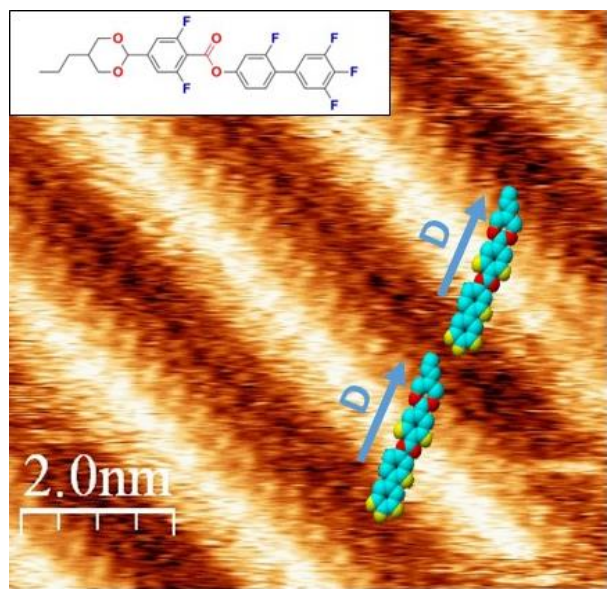

b

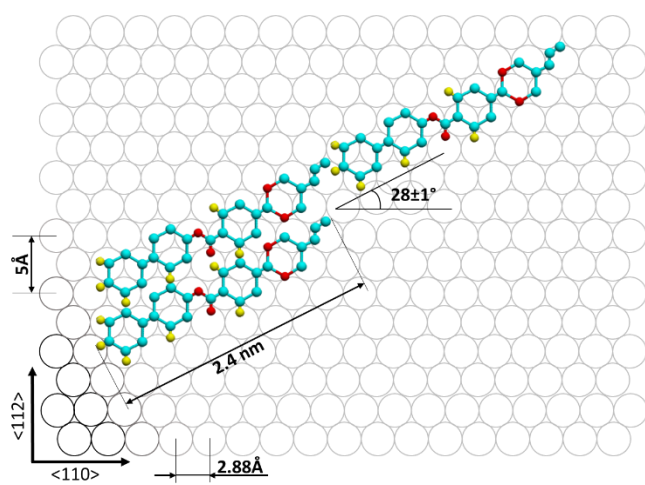

c

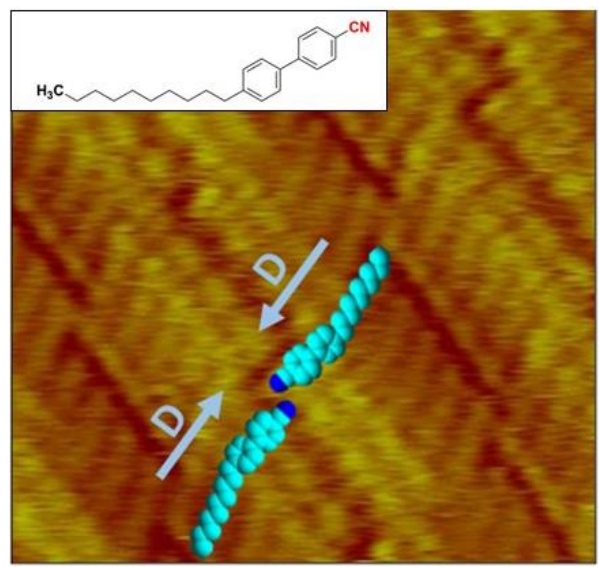

d

Figure 1. a) Large-scale STM image of a monomolecular DIO film on Au(111) surface. $I_t = $ 200 pA, $U_t = $ 250 mV. b) High-resolution image of adsorbed DIO molecules. The schematic position of adsorbed molecules is illustrated with a model

drawing. The molecular dipoles **D** align unidirectionally. $I_t = 200$ pA, $U_t = 250$ mV. c) Schematic arrangement of DIO molecules on Au(111) surface with the optimized energy of interaction with the Au lattice. d) Self-organized monolayer of mesogenic 10CB molecules on Au(111) surface with a head-to-head packing. $I_t = 300$ pA, $U_t = 300$ mV.

The STM image in Fig. 1b resolves individual DIO molecules within the layers. Conjugated π-electron systems are known to provide a strong STM contrast [29,30], thus the bright region within each molecule can be identified with its aromatic core. Darker regions represent lower electron density associated with alkyl chains. From Fig. 1b, the molecular length is estimated as $l = (24.0 \pm 0.2)$ Å, and the side-by-side molecular separation as $w = (4.4 \pm 0.2)$Å, somewhat larger than 4.2 Å reported for the bulk DIO in the SmZ$_A$ phase [7]. The larger separation $w$ is expected since the molecules lie flat at the substrate. The width of a molecular row in Fig. 1a is $d = (21.0 \pm 0.2)$Å. The tilt angle of the molecules away from the layers' normal is then $\alpha = \arccos(d/l) = (28 \pm 1)°$. The tilt implies that the neighboring molecules are shifted with respect to each other, by $\arcsin(d/l) \approx (2.3 \pm 0.2)$ Å. The DIO monolayer structure in Fig. 1a,b remains the same when the preparation temperature varies in the range $(20 - 80)°$C. Remarkably, the STM images show no molecular dimerization or overlapping.

To quantify the contrasts in the STM images, quantum-chemical calculations are performed for a neutral isolated DIO molecule (Supplemental Material [31], see also references [2,32–34] therein). The calculations show that estimated length and width of the backbone and the shapes of frontier molecular wavefunctions defined by a π-conjugated core of the molecule are in good agreement with the bright regions in STM images, Fig. 1b. The distribution of contrasts confirms the polar orientational order of ferronematic molecules as depicted in Fig. 1c: the DIO electric dipoles (~9.43 D [2]) point in the same direction.

The observed polar ordering of DIO molecules differs drastically from arrangements in monolayers of conventional liquid crystals, such as cyanobiphenyls [35,36] and antiferroelectric smectics [37,38], in which the electric dipoles of neighboring molecules are antiparallel. To further demonstrate the difference, we explored monolayers of 4,4'-n-decyl-cyanobiphenyl (10CB) at the same Au(111) substrates. In its bulk form, 10CB shows a direct transition from the isotropic to the smectic-A phase, Iso – 50.2 °C – SmA – 44 °C – Cr. The STM textures reveal head-to-head pairing of molecules, Fig. 1d, with their dipoles **D**, on the order of 6 Debye, being antiparallel. This result is in line with other STM observations of apolar orientations in monolayers of molecules which form apolar bulk LC phases [35,36,39-46].

The smectic C-like polar structure in Fig.1b,c is different from the bulk 3D arrangements of DIO, which form the paraelectric N, antiferroelectric $SmZ_A$ and ferroelectric $N_F$ phases [2,7]. The observed polar order and shifts along the polar axis cannot be explained by a simple model of point-like dipoles. Instead, the data provide support for the model of longitudinal surface charge density wave proposed by Madhusudana [47]. The $N_F$ molecules are modeled as cylindrical rods with a chain of alternating positive and negative electric charges along the long axes. A shift of one molecule with respect to its side neighbor by half-period $\zeta$ of the charged wave produces strong electrostatic attraction of two molecules oriented in a syn-polar fashion. This polar arrangement is of a lower electrostatic energy than the antiparallel assembly when the distances between the molecules are short [47]. As the separations increase, the contributions from different charges overlap and an antiparallel arrangement prevails.

Figure 2a presents the static site charge distribution of a DIO molecule, calculated by DFT with the hybrid B3LYP functional and the split valence basis set 6-31G* using the Gaussian'03 package [34]. The molecule can be conditionally split into parts along its length, with the charges summed up within each part. One possible summation scheme with six parts is shown in Fig. 2a. The wave of alternating positive and negative charges is similar to the Madhusudana's model [47]. Therefore, the shift of a DIO molecule with respect to its neighbor can result from the electrostatic attraction of oppositely charged groups of the two molecules, as schematized in Fig. 2b. An important condition of the model [47] is that the end charges are relatively weak, which is the case of the DIO molecule, one end of which contains three fluorine substitutions in ortho positions and the other end is a short aliphatic chain, Fig. 2.

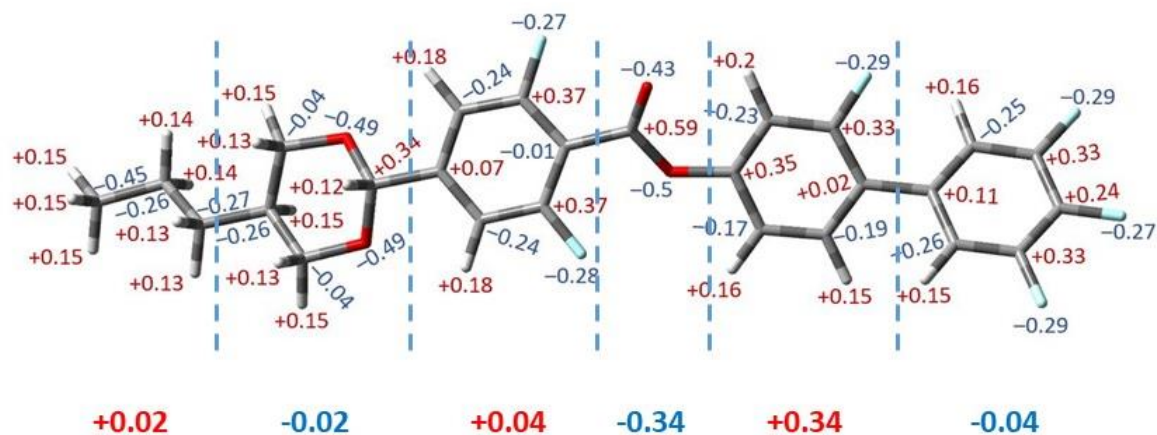

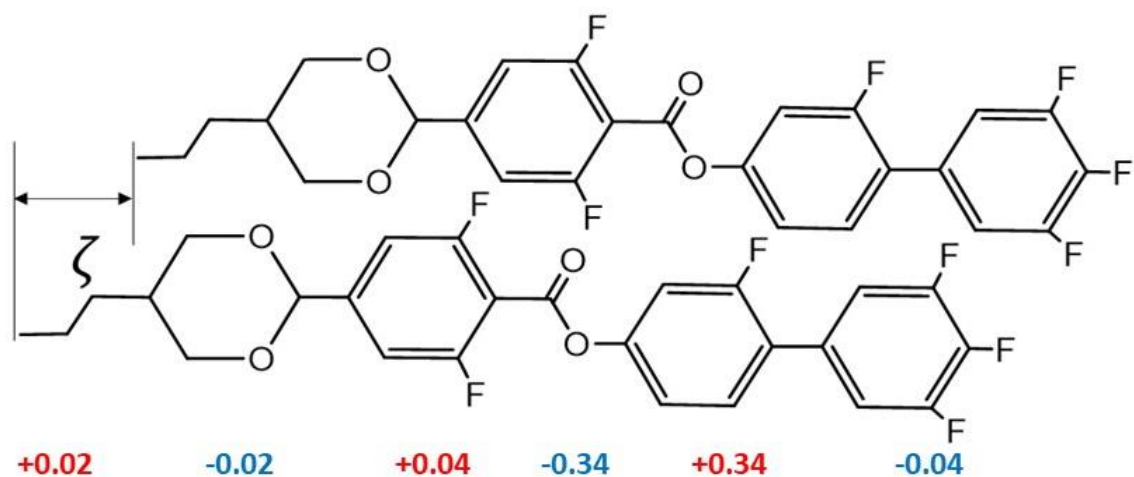

Figure 2. DIO molecules: (a) electric charge distribution evaluated using the Gaussian'03 package; red and blue are positive and negative charges, respectively; (b) scheme of the electric charge longitudinal wave and shifts of DIO molecules providing side-by-side attractions of the molecules in syn-polar orientation.

**Conclusion**

We presented the first direct observation of polar molecular ordering in the monolayer of ferroelectric nematic mesogens, which is in strike contrast to the behavior of conventional nematics. Molecules of conventional nematics adsorbed on atomically flat crystal surfaces exhibit non-polar ordering, even if they carry a significant electric dipole moment, as in the case of cyanobiphenyls, Fig.1d. The molecules of ferroelectric nematic DIO, however, adsorbed on a reconstructed Au(111) surface exhibit a high degree of polar ordering, Fig. 1b. The DIO molecules shift with respect to each other along the alignment axis so that the 2D monolayer resembles the smectic C, in which molecules within the layers are tilted. The experiments support the model [47] of ferroelectric nematic ordering in which the syn-polar molecules electrostatically attract each other in side-by-side positions thanks to the wave of alternating positive and negative electric charges along t the molecules. Note that the waves of alternating charges along the molecular axes allows electrostatic attraction of similarly charged biomolecules such as nucleic acids in an ionic environment [48].

The polarly ordered $N_F$ monolayers exhibit an extremely thin form factor, which makes them useful for organic microelectronic and nanoelectronic applications. Overall, self-organization is a key process in many scientific and technological fields, and understanding how ferroelectric nematic molecules organize themselves on surfaces is essential for advancing general knowledge and developing new materials.

**Acknowledgments**

The quantum-chemical calculations were run on the computer cluster at Bogolyubov Institute for Theoretical Physics of NASU. This work was supported by NASU projects No. 0121U109816, 0123U100832; the Long-term program of support of the Ukrainian research teams at the PAS Polish Academy of Sciences carried out in collaboration with the U.S. National Academy of Sciences with the financial support of external partners via the agreement No. PAN.BFB.S.BWZ.356.022.2023; the NATO Science for Peace and Security Programme grant SPS G6030; Kent State University's Ukraine Scholars Fund. ODL acknowledges the support of NSF grant ECCS-2122399. YYL acknowledges the European fund for displaced scientists (EFDS program from ALLEA and Breakthrough Prize Foundation) and MSCA4Ukraine fellowship scheme (No. 1232376), which is funded by the European Union.

# SUPPLEMENTARY INFORMATION

To quantify the contrasts in the STM images, quantum-chemical calculations are performed for a neutral isolated DIO molecule. The geometry optimization and electronic structure calculations are carried out at the density functional theory (DFT) level with the hybrid B3LYP functional [32,33] and the split valence basis set 6-31G* using Gaussian'03 program package [34]. The calculations show that the optimized ground state is the *trans* conformation of the molecular backbone, Fig. S1. The alternative *cis* conformation is not energetically favorable. The *cis*-conformation implies a "bent" shape (almost 90 degrees), which is not observed in the STM experiment. The calculated dipole of the DIO molecules is 9.42 Debye (9.43 [2]). The estimated length and width of the backbone and the shapes of frontier molecular wavefunctions defined by a π-conjugated core of the molecule are in good agreement with the bright contrast regions in STM images, Fig. 1b. Since the calculated HOMO-LUMO gap for the DIO molecule exceeds 4eV, which is substantially larger than the applied bias tunneling voltages (50-500mV), there is no dependence of STM contrast on the applied voltage. The distribution of contrasts observed in the STM image suggests the polar orientational order of ferronematic molecules at the Au(111) substrate as depicted in Fig. 1c: permanent electric dipoles of neighboring molecules are pointed in the same direction giving rise to the SAM with a polar ordering.

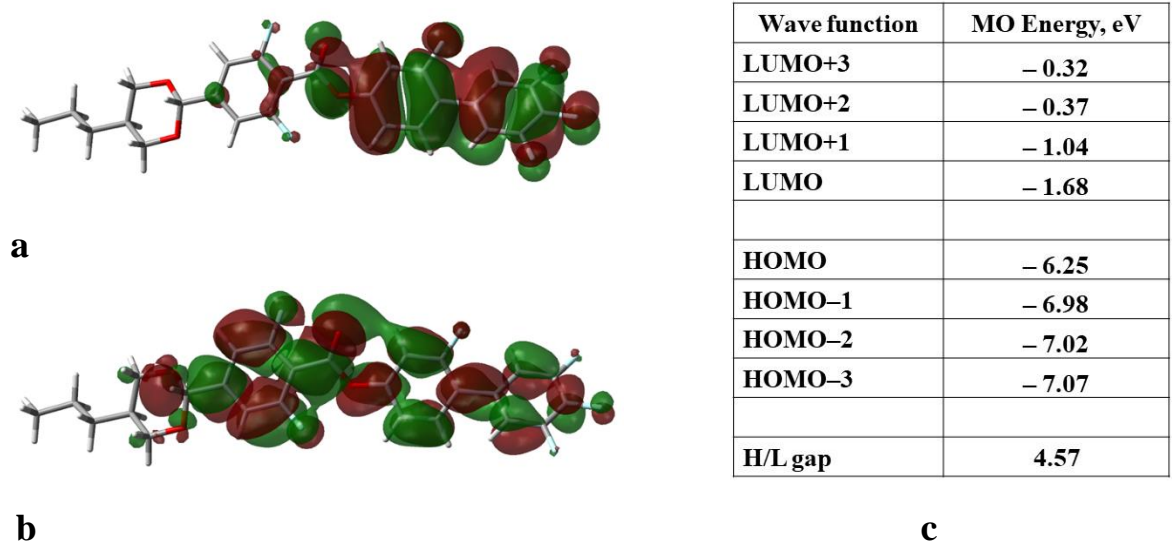

| Wave function | MO Energy, eV |
|---|---|
| LUMO+3 | – 0.32 |
| LUMO+2 | – 0.37 |
| LUMO+1 | – 1.04 |
| LUMO | – 1.68 |
|  |  |
| HOMO | – 6.25 |
| HOMO–1 | – 6.98 |
| HOMO–2 | – 7.02 |
| HOMO–3 | – 7.07 |
|  |  |
| H/L gap | 4.57 |

a

b

c

Figure S1. The profiles of calculated (a) HOMO (–6.25 eV) and (b) LUMO (–1.68 eV) wavefunctions in the neutral ground state of the molecule. Different colors of the molecular orbital surfaces represent a phase of the wave functions (red for positive, green for negative). (c) The energies of four frontier HOMOs and LUMOs levels for neutral molecules in the gas phase.